\documentclass[sigconf]{acmart}

\usepackage{hyperref}
\usepackage{multirow}
\usepackage{subcaption}
\usepackage{caption}
\usepackage{setspace}
\usepackage{amsmath}
\usepackage[english]{babel}
\definecolor{mygray}{gray}{0.5}
\usepackage{float}
\usepackage{flushend}
\usepackage{mathtools}
\usepackage{soul}
\usepackage{arydshln}
\usepackage{algpseudocode}
\usepackage{algorithm}
\usepackage{url}
\usepackage{xcolor}
\usepackage{graphicx}

\usepackage{enumitem}

\long\def\comment#1{}
\long\def\comments#1{}

\author{Parker Carlson}
\affiliation{%
  \institution{University of California, Santa Barbara}
  \city{Santa Barbara}
  \state{California}
  \postcode{93106}
  \country{USA}
}
\author{Wentai Xie}
\affiliation{%
  \institution{University of California, Santa Barbara}
  \city{Santa Barbara}
  \state{California}
  \postcode{93106}
  \country{USA}
}
\author{Shanxiu He}
\affiliation{%
  \institution{University of California, Santa Barbara}
  \city{Santa Barbara}
  \state{California}
  \postcode{93106}
  \country{USA}
}
\author{Tao Yang}
\affiliation{%
  \institution{University of California, Santa Barbara}
  \city{Santa Barbara}
  \state{California}
  \postcode{93106}
  \country{USA}
}
\renewcommand{\shortauthors}{Parker Carlson, Wentai Xie, Shanxiu He, and Tao Yang}

\copyrightyear{2025}
\acmYear{2025}
\setcopyright{cc}
\setcctype{by}
\acmConference[SIGIR '25]{Proceedings of the 48th International ACM SIGIR Conference on Research and Development in Information Retrieval}{July 13--18, 2025}{Padua, Italy}
\acmBooktitle{Proceedings of the 48th International ACM SIGIR Conference on Research and Development in Information Retrieval (SIGIR '25), July 13--18, 2025, Padua, Italy}\acmDOI{10.1145/3726302.3730183}
\acmISBN{979-8-4007-1592-1/2025/07}

\makeatletter
\gdef\@copyrightpermission{
 \begin{minipage}{0.3\columnwidth}
  \href{https://creativecommons.org/licenses/by/4.0/}{\includegraphics[width=0.85\textwidth]{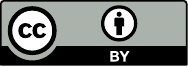}}
 \end{minipage}\hfill
 \begin{minipage}{0.7\columnwidth}
  \href{https://creativecommons.org/licenses/by/4.0/}{This work is licensed under a Creative Commons Attribution International 4.0 License.}
 \end{minipage}
 \vspace{5pt}
}
\makeatother

\begin{document}

\title{Dynamic Superblock Pruning for Fast Learned Sparse Retrieval}

\begin{abstract}

This paper proposes superblock pruning (SP) during top-$k$ online document retrieval
for learned sparse representations.
SP  structures  the sparse index as a set of superblocks on a sequence of document blocks
and conducts a superblock-level  selection to decide if
some superblocks  can be pruned before visiting their child blocks.
SP generalizes  the previous flat block or cluster-based pruning, 
allowing the early detection of groups of documents that cannot or are less likely to
appear in the final top-$k$ list.
SP can accelerate sparse retrieval 
in a rank-safe or approximate  manner under a high-relevance competitiveness constraint.
Our experiments show that the proposed scheme significantly outperforms state-of-the-art baselines
on MS MARCO passages on a single-threaded CPU.

\end{abstract}



\maketitle

\section{Introduction}

Sparse retrieval models such as BM25
and learned sparse representations 
~\cite{Mallia2021deepimpact, Lin2021unicoil, Formal2021SPLADE, shen2023lexmae}
are popular for inexpensive CPU-only servers, 
 since they can take advantage of fast inverted index implementations. 
A traditional speed optimization for sparse retrieval is dynamic rank-safe index pruning,
which accurately skips the evaluation of low-scoring documents that are unable to appear in the final top-$k$ results~\cite{Turtle1995,2013CandidateFiltering, WAND, BMW, 2021WSDMliveBlock}.
These methods have been extended to unsafe pruning (approximate search),
with early work including threshold overestimation~\cite{2012SIGIR-SafeThreshold-Macdonald, 2013WSDM-SafeThreshold-Macdonald, 2017WSDM-DAAT-SAAT}
and early termination~\cite{2015ICTIR-anytime-ranking-Lin,2022IOQP}.
Recent work in dynamic pruning includes block-based retrieval,
where documents are assigned to blocks (clusters),
and block-level information is used to improve index-traversal order and prune
groups of low-scoring documents~\cite{2022ACMTransAnytime,mallia2024BMP,qiao2024ASC,2024SIGIR-SparseApproximate}.

This paper expands upon previous work in block-based pruning for both safe and approximate settings.
We introduce Superblock  Pruning (SP),
that uniformly aggregates a sequence of document blocks into a superblock and conducts
dynamic superblock-level pruning.
This gives SP more opportunities to skip document blocks and  
accelerate retrieval in a rank-safe or probabilistically rank-safe manner.
Pruning a superblock avoids both calculating subblock maximum scores and scoring 
documents within its subblocks.  
Our design assigns a constant number of document blocks to each superblock
to simplify vectorization,
cache optimization, and to provide two-level pruning with a probabilistic safeness guarantee.

Our evaluation shows that under a high-relevance budget requirement, 
SP is significantly faster than the other state-of-the-art 
baselines BMP, ASC, and Seismic~\cite{mallia2024BMP,qiao2024ASC,2024SIGIR-SparseApproximate}
for SPLADE~\cite{Formal2021SPLADE,Formal_etal_SIGIR2022_splade++} 
and E-SPLADE~\cite{lassance2022efficiency}
on MS MARCO Passage ranking.

\vspace{-3mm}
\section{Background and Related Work}
\label{sect:background}

{\bf Problem definition.}
Sparse document retrieval identifies top-$k$ ranked candidates that  match a query.
Each document in a collection is modeled as a sparse vector.
These candidates are ranked using a
simple formula, where the rank score of each document $d$ is defined as:
$RankScore(d) =
\sum_{t \in Q} q_t \cdot w_{t,d}$,
where $Q$ is the set  of search terms in the given query,
$w_{t,d}$ is a weight contribution of  term $t$  in document $d$, scaled by a corresponding query term weight $q_t$.
Term weights can be based on a lexical model  such as BM25~\cite{Jones2000}
or are learned from a neural model.
For sparse representations, retrieval algorithms typically use
an \textit{inverted index},
though recent work has explored the usage of forward and hybrid indexes~\cite{mallia2024BMP,2024SIGIR-SparseApproximate}.

\noindent
{\bf Threshold-based skipping.} 
During sparse retrieval,
a pruning  strategy computes the upper bound rank score of a candidate document $d$,
referred to as $Bound(d)$.
If $Bound(d) \leq  \theta$, where $\theta$ is the heap threshold to enter the top-$k$ list,
this document can be safely skipped.
A  retrieval method is  called {\em rank-safe} if it  guarantees that the top-$k$ documents returned are the $k$ highest scoring documents.
WAND~\cite{WAND} 
uses the maximum term weight of documents in a posting list  for their score upper bound,
while BMW~\cite{BMW} 
and its variants (e.g. VBMW~\cite{Mallia2017VBMW}) 
use  block-based maximum weights. 
MaxScore~\cite{Turtle1995} 
uses  a similar skipping strategy with
term partitioning.
Live block filtering
~\cite{2013CandidateFiltering, 2021WSDMliveBlock} 
clusters  document IDs within a range and estimates  a range-based max score  for  pruning.
The above methods are all rank-safe.
Threshold estimation~\cite{2020CIKMComparison,2015BMWheuristics,2019SIGIR-Petri,2018SIGIRKane}
predicts the final threshold value (safely or unsafely)
and accelerates early query processing.
Threshold overestimation is a common approximate strategy that deliberately 
overestimates the current top-$k$ threshold 
by a factor~\cite{
2012SIGIR-SafeThreshold-Macdonald, 2013WSDM-SafeThreshold-Macdonald, 2017WSDM-DAAT-SAAT}.

\noindent
{\bf Block or cluster based pruning.}
Block based skipping~\cite{2013CandidateFiltering, 2021WSDMliveBlock, mallia2024BMP}
divides  documents into blocks to estimate the block-wise maximum rank score for pruning.
Often, documents are reordered before blocking using a Bipartite Partitioning algorithm~\cite{
    dhulipala2016bp, mackenzie2019bp} that groups similar documents together.
Conceptually, a block is the same as cluster-based 
skipping~\cite{2004InfoJ-ClusterRetr,2017ECIR-SelectiveSearch, 2022ACMTransAnytime}. 
Representative recent studies~\cite{2022ACMTransAnytime, mallia2024BMP,2024CIKM-SeismicWave}
order the visitation of the blocks by their maximum rank score. 
ASC~\cite{qiao2024ASC}
extends the above cluster-based pruning studies
by introducing probabilistic rank-safeness
which increases index-skipping opportunities while maintaining competitive relevance.
BMP~\cite{mallia2024BMP} optimizes execution with quantization, 
SIMD $BoundSum$ computation, 
partial block sorting, and query pruning.
The main optimization in Seismic~\cite{2024SIGIR-SparseApproximate} 
is aggressive static inverted index pruning while fully scoring documents with an unpruned forward index.
Like BMP~\cite{mallia2024BMP}, Seismic also incorporates threshold overestimation, query pruning,
and dynamic cluster (block) maximum pruning. 
SP incorporates BMP's optimizations and operates on a given static index;
our evaluation does not use static pruning.

\noindent
{\bf Other efficiency optimization techniques.}
There are orthogonal techniques to accelerate learned sparse retrieval.
BM25-guided pruning skips documents during index traversal~\cite{mallia2022faster,20232GT}.
Static index pruning~\cite{2023SIGIR-Qiao,2023SIGIR-SPLADE-pruning}
removes  low-scoring  term weights during index generation.
An efficient version of SPLADE~\cite{lassance2022efficiency} uses
L1 regularization for query vectors, and dual document and query encoders.
Term impact decomposition~\cite{mackenzie2022accelerating} partitions each posting list 
into two groups 
with high and low impact weights.  
Our work is complementary to the above techniques.

\vspace{-5pt}

\section{Dynamic Superblock Pruning}
\setlength{\belowdisplayskip}{4pt} \setlength{\belowdisplayshortskip}{4pt}
\setlength{\abovedisplayskip}{4pt} \setlength{\abovedisplayshortskip}{4pt}
We start from a flat block-based index
approach ~\cite{mallia2024BMP, 2022ACMTransAnytime, 2024SIGIR-SparseApproximate,qiao2024ASC},
where a document collection is divided into a sequence of $N$ blocks $\{B_1, \cdots, B_N \}$.
Like BMP, we assume that each block uniformly contains $b$ documents. 
Like previous work~\cite{mallia2024BMP, 2022ACMTransAnytime},
blocks are visited in decreasing order of $BoundSum$ values.
\begin{equation}
\label{eq:boundsum}
	BoundSum(B_i) = \sum_{t \in Q}  \max_{d \in B_i} q_t \cdot w_{t,d}.
\end{equation}
The visitation to block $B_i$ can be safely pruned if $ BoundSum(B_i) \leq \theta $,
where $\theta$ is the current top-$k$ threshold. 
If this block is not pruned, then document-level index traversal can be 
conducted within each block following a standard retrieval algorithm.

\begin{figure}[tbhp]
\begin{center}
  \includegraphics[width=\columnwidth,height=0.3\columnwidth]{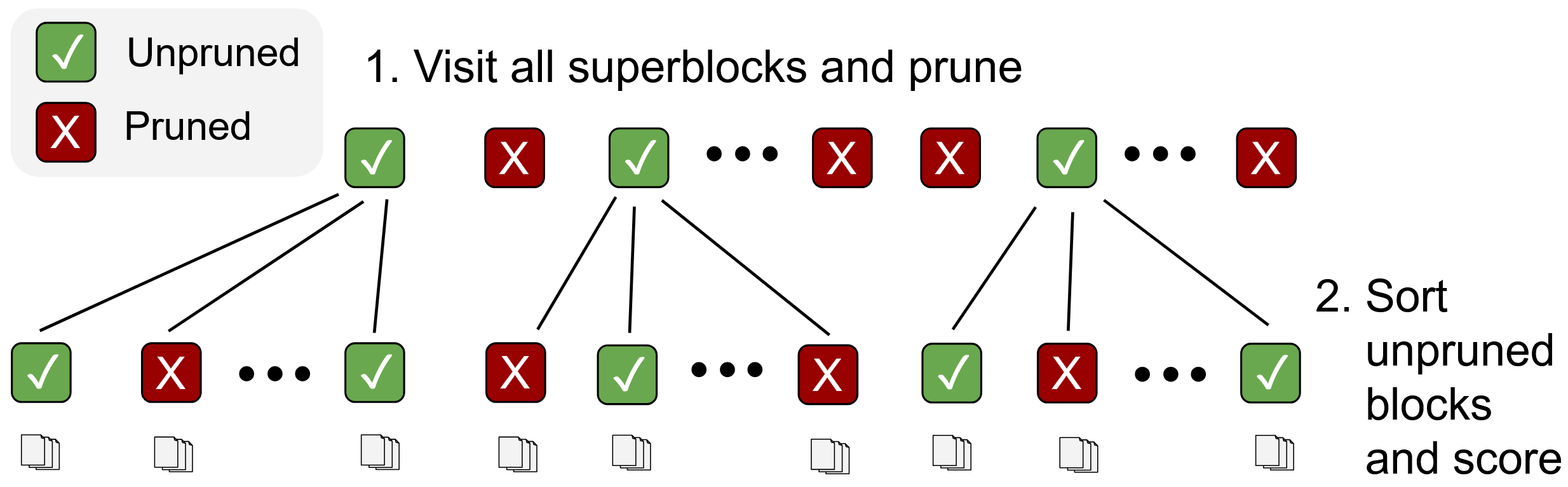}
\end{center}
  \caption{Superblock and block pruning during traversal}
  \label{fig:SPtrees}
\end{figure}

We propose to uniformly aggregate a sequence of $c$ consecutive document blocks
into one superblock, 
and then 
conduct online index traversal in a top-down manner as illustrated in  
Figure~\ref{fig:SPtrees}.
We assume the documents are reordered based on a similarity-based clustering strategy
like Bipartite Partitioning~\cite{dhulipala2016bp, mackenzie2019bp} used in BMP.

During offline indexing, 
we precompute
the maximum term weight for each term $t$ from documents contained within
each block and superblock. For superblocks, we also compute the average maximum term weight.
Specifically, given document block $B$, $W_{B,t}  = \max_{d \in B}  w_{t,d}.$
Given superblock  $X$ with $c$ child blocks $\{B_{1}, \cdots, B_{c}\}$,
$W_{X,t}  = \max_{B_{i} \in X}  W_{B_i,t}; \ \ \overline{W}_{X,t}  = \frac{1}{c} \sum_{B_{i} \in X}  W_{B_i,t}.$
\comments{ %
\[
W_{B,t}  = \max_{d \in B}  w_{t,d}.
\]
Given superblock  $X$ with $c$ child blocks $\{B_{1}, \cdots, B_{c}\}$,
\[
W_{X,t}  = \max_{B_{i} \in X}  W_{B_i,t}; \ \ \overline{W}_{X,t}  = \frac{1}{c} \sum_{B_{i} \in X}  W_{B_i,t}.
\]
}

Online query inference begins by computing bound information  of all superblocks and pruning them,
then descends to compute bounds for  blocks and prune them. 
Specifically, SP conducts the following dynamic pruning steps:

\begin{itemize}[leftmargin=*]
\item
Given superblock $X$, we compute the maximum and average 
rank score bound  
of documents within this  superblock as follows:
\begin{equation}
\label{eq:subtreemax}
SBMax(X) = \sum_{t \in Q}   q_t\cdot W_{X,t};  \ 
\overline{SBMax}(X) =
\sum_{t \in Q} q_t \cdot \overline{W}_{X,t}.
\end{equation}

Let $\theta$ be the current top-$k$ retrieval threshold for handling query $Q$.
Any superblock $X$ is pruned when its maximum and average superblock bounds satisfy
$SBMax(X)  \le \frac{\theta}{\mu} \mbox{ and } \overline{SBMax}(X) \le \frac{\theta}{\eta}$
where parameters $\mu$ and $\eta$ satisfy $0<\mu \leq \eta \le 1$.

\item Given a document block B, we prune $B$  if 
$BoundSum(B)  \le \frac{\theta}{\eta}.$
\item
For all un-pruned blocks,
the corresponding document blocks are sorted and scored in a descending order of their 
$BoundSum$ values,
and a standard retrieval algorithm is applied to score documents within each block.
BMP uses a forward-index approach which is fast for small block sizes,  
and we adopt the same strategy.

\end{itemize}

The  two-parameter pruning setup is  inspired by ASC~\cite{qiao2024ASC}, with two substantial differences.
First, ASC requires a random partitioning within each block to ensure probabilistic safeness while
we build a superblock from consecutive blocks without randomness.
Second, ASC computes a tighter block bound by scoring multiple segments per block during online search,
whereas we only compute a single bound per block.
Specifically, given $n$ segments within  block $B_i$, ASC computes its bound as
$MaxSBound(B_i) = \max_{j=1}^{n} \sum_{t\in Q} q_t \cdot  \max_{d\in S_{i,j}} w_{t,d}$
while we compute the superblock maximum rank bound as 
$ SBMax(X) = \sum_{t \in Q}   q_t\cdot \max_{B_i \in X}   \max_{d \in B}  w_{t,d}.$
Our sum over the query terms is outside both maxes,
leading to a looser superblock bound compared to ASC,
but with the advantage that $\max_{B_i \in X}   \max_{d \in B}  w_{t,d}$ is computed offline,
reducing block filtering overhead for a large number of blocks.
Moreover,
SP makes up for this looser bound
because SP also prunes at the block level,
where bounds are inherently tight by nature of a small block size.

\noindent
{\bf CPU cache usage for score bounding}
We compute Formulas~(\ref{eq:boundsum}) and (\ref{eq:subtreemax}) using SIMD instructions.
When computing either of these formulas sequentially for all query terms without block skipping,
modern compilers can easily vectorize their implementation,
and modern CPUs can effectively prefetch their data.
However, 
compilers struggle to optimize the block-level bound computation
because of the irregular and non-consecutive data access from superblock pruning.
Thus, SP needs to explicitly control the CPU cache reuse pattern
in computing block-level bounds. %

\begin{figure}[thb]
    
    \raggedright{ {\bf Option 1:} Term-at-a-time filtered $BoundSum$ computation}
    \centering
    \resizebox{0.9\columnwidth}{!}{
    \fbox{
        \begin{minipage}{.45\textwidth}
    \begin{algorithmic}
    \For{term  $\in$ query}
        
        \For{every unpruned superblock $s$}
            \State Accumulate $BoundSum$ for all blocks of $s$ 
        \EndFor
    \EndFor
    \end{algorithmic}
        \end{minipage}
    
    }
    }
    
    \raggedright{{\bf Option 2:} Superblock-at-a-time filtered $BoundSum$ computation}
    \centering
    \resizebox{0.9\columnwidth}{!}{
    \fbox{
        \begin{minipage}{.45\textwidth}
    \begin{algorithmic}
    \For{every unpruned superblock $s$}
        
        \For{term $\in$ query}
            \State Accumulate $BoundSum$ for all blocks of $s$ 
        \EndFor
    \EndFor
    
    \end{algorithmic}
    \end{minipage}
    }
    }

   \vspace*{-0.5em} 
    \caption{Control flow for maximum score computation}
    \label{fig:cache}
   \vspace*{-0.5em} 
    \end{figure}

Figure~\ref{fig:cache} shows two control flow options for
calculating the filtered $BoundSum$ of Formula~(\ref{eq:boundsum})
with different CPU cache access patterns.
Option 1 conducts term-at-a-time accumulation,
where the $BoundSum$ value for all unpruned blocks is accumulated for each term in sequence.
Option 2 conducts superblock-at-a-time accumulation,
where the document blocks within each superblock are fully scored for all terms before proceeding to the next unpruned superblock.
Option 2 allows the accumulation registers for the final result to be reused in
the inner loop and obtains better L1 cache performance. 
SP adopts Option 2, and 
Section~\ref{sect:eval} shows that Option 2 is up to 1.89x faster than Option 1 in our tested scenario.

\noindent
{\bf Rank-safeness properties.}
\label{sect:safeness}
SP has  a rank-safe $\mu$-competitiveness property like ASC.
Define $Avg(x, A)$ as the average rank score of the top-$x$ results by algorithm $A$.
Let an integer $k' \leq  k$.
We can prove that the average top-$k'$ rank score of SP  
is the same as any rank-safe retrieval algorithm $R$ within a factor of $\mu$.
Namely, $Avg(k', {\rm SP }) \ge \mu Avg(k',  R)$.
As an extra safeguard, SP provides probabilistic safeness
if we can assume the rank scores  of documents are independently and identically distributed within each superblock. With this assumption, for any superblock $X$ pruned by SP,
the pruned  document $d$ within $X$  satisfies:
\vspace*{-0.4em}
\begin{small}
     \begin{multline*}
          E[RankScore(d)] 
          = \frac{1}{b \cdot c} \sum_{z \in X} RankScore(z) \\
     \leq   \frac{1}{c}\sum_{t \in Q}  \sum_{B_i \in X} q_t \cdot W_{B_i,t}
     =  \overline{SBMax}(X) \leq \frac{\theta_{{\rm SP}}}{\eta} 
     \end{multline*}
\end{small}
\vspace*{-0.4em}

where $\theta_{{\rm SP}}$ is the top-$k$ threshold of SP during above pruning.
Then following~\cite{qiao2024ASC}, we can show that with $k' \leq k$,
the average top-$k'$ rank score of SP is within the expected value of
any rank-safe retrieval algorithm $R$ by a factor of $\eta$.
Namely,  $E[Avg(k', {\rm SP })] \ge \eta E[Avg(k',  R)]$ 
where $E[\cdot]$ denotes the expected value.

\noindent
{\bf  Extra space cost for superblock  pruning.}
Compared to BMP, the extra space cost in SP 
is to maintain maximum and average term weights for each superblock.
Given $N$ document blocks, the number of superblocks is $\lceil\frac{N}{c}\rceil$.
In our evaluation with MS MARCO, 
$c=64$, $b=8$, $N\approx 1.1M$. If $b=16$, $N \approx 0.55M$. 
Each superblock max score is quantized to 8 bits and each average to 16 bits.
This results in about 2GB of extra space with $b=8$ and 1GB for $b=16$.

Section~\ref{sect:eval} follows 
Seismic~\cite{2024SIGIR-SparseApproximate}
to report the latency when the corresponding sparse index is uncompressed in memory during query 
processing  for BMP, SP, and Seismic.  
For MS MARCO passages, BMP's uncompressed raw index is up to 37GB while SP's maximum index size is 39GB.
Seismic's uncompressed index is smaller at 13GB because it uses static index pruning.
For ASC, we report latency using its compressed index with a total size of 6.2GB
because its PISA base~\cite{mallia2019pisa} has fully optimized index decompression.

\section{Experimental Studies}
\label{sect:eval}

\begin{table}[htbp]
    \small
        \centering
        \setlength{\belowcaptionskip}{-2pt}
        \setlength{\abovecaptionskip}{8pt}
        \caption{Mean response time ($ms$) and mean reciprocal rank (MRR@10) at a fixed Recall@$k$ budget for SPLADE}
        \renewcommand{\scriptsize}{\small}
    \label{tab:main}
    \resizebox{\columnwidth}{!}{
        \begin{tabular}{l|cccccccc}
        \hline
        \hline
        \textbf{Recall} & \multicolumn{2}{c}{\textbf{99\%}} & \multicolumn{2}{c}{\textbf{99.5\%}} & \multicolumn{2}{c}{\textbf{99.9\%}} & \multicolumn{2}{c}{\textbf{Rank-Safe}} \\
        \textbf{Budget} & MRT & MRR & MRT & MRR & MRT & MRR & MRT & MRR \\ 
        \hline
        \multicolumn{9}{c}{$k$=10} \\  
        \hline

        MaxScore & -- & -- & -- & -- & -- & -- & 75.7 {\scriptsize (35x)} & 38.1 \\
        ASC & 4.70 {\scriptsize (7.5x)} & 37.9 & 5.59 {\scriptsize (7.8x)} & 38.1 & 6.44 {\scriptsize (8.2x)} & 38.1 & 7.19 {\scriptsize (3.3x)} & 38.1 \\ %
        Seismic & 2.06 {\scriptsize (3.3x)} & 38.1 & 2.57 {\scriptsize (3.6x)} & 38.2 & 3.01 {\scriptsize (3.8x)} & 38.4 & -- & -- \\
        BMP & 1.44 {\scriptsize (2.3x)} & 38.1 & 1.49 {\scriptsize (2.1x)} & 38.1 & 1.88 {\scriptsize (2.4x)} & 38.2 & 2.70 {\scriptsize (1.3x)} & 38.1 \\ %
        SP & \textbf{0.629} & 37.7 & \textbf{0.715} & 37.9 & \textbf{0.785} & 38.1 & \textbf{2.15} & 38.1 \\
        \hline
        \multicolumn{9}{c}{$k$=1000} \\
        \hline
        MaxScore & -- & -- & -- & -- & -- & -- & 124 {\scriptsize (12x)} & 38.1 \\
        ASC & 15.8 {\scriptsize (9.1x)} & 38.1 & 18.9 {\scriptsize (9.4x)} & 38.1 & 25.4 {\scriptsize (5.5x)} & 38.1 & 33.5 {\scriptsize (3.2x)} & 38.1 \\ %
        Seismic & 5.72 {\scriptsize (3.3x)} & 38.3 & 7.18 {\scriptsize (3.6x)} & 38.4 & 10.5 {\scriptsize (2.3x)} & 38.4 & -- & -- \\
        BMP  & 4.99 {\scriptsize (2.9x)} & 38.2 & 5.25 {\scriptsize (2.6x)} & 38.2 & 7.26 {\scriptsize (1.6x)} & 38.2 & 13.9 {\scriptsize (1.3x)} & 38.1 \\
        SP & \textbf{1.74} & 37.9 & \textbf{2.01} & 37.9 & \textbf{4.64} & 38.2 & \textbf{10.5} & 38.1 \\ 
        \hline
        \hline
    
        \end{tabular}
    }
\end{table}

We evaluate on the MS MARCO Passage ranking dataset~\cite{2016MSMARCO} with 8.8 million English passages.
We use the standard metrics of mean reciprocal rank (MRR@10) and recall at 
positions 1000 (when $k=1000$) or 10 (when $k=10$)
for the Dev queries, and nDCG@10 for the TREC Deep Learning (DL) 2019 and 2020 queries. 
We run all experiments using a single thread on a Linux system
with an Intel i7-1260P, 64GB of RAM, and AVX2 instructions.
SP is compiled using rustc 1.84 with -O3 optimization.
We preload the index into memory,
and in following common timing practice of using a "warm" index,
we run search five times, drop the first two runs, and report latency as the average of the remaining runs.

We compare SP against three state-of-the-art block-based retrieval algorithms: 
BMP~\cite{mallia2024BMP}, Seismic~\cite{2024SIGIR-SparseApproximate}, and ASC~\cite{qiao2024ASC}.
We also compare against PISA's~\cite{mallia2019pisa} implementation of MaxScore~\cite{Turtle1995}.
We do not compare against Seismic-Wave~\cite{2024CIKM-SeismicWave};
its use of a corpus neighbor proximity graph is an orthogonal optimization that can be applied to any method.
We test these methods on two learned sparse retrieval methods,
SPLADE~\cite{Formal_etal_SIGIR2022_splade++} and Efficient-SPLADE~\cite{lassance2022efficiency}.
We run all algorithms using their official code release;
our code is available at \url{https://github.com/thefxperson/hierarchical_pruning}.

\noindent
{\bf Baseline Comparison on SPLADE.}
Table~\ref{tab:main}
presents an overall comparison of these methods under a tight relevance budget. 
Following~\cite{Big-ANN},
"recall budget" indicates the percentage of preserved recall relative to safe search;
for instance, if safe search achieves a recall of 98.36 for $k$=1000,
then a 99\% recall budget represents the fastest time that 
a method can achieve a recall of at least 97.38.
Notice this is a ratio of recall,
not the degree of overlap of the results.
We report mean latency in milliseconds,
and speedup relative to the fastest method in parenthesis.
For each algorithm, we start from the published best parameters then vary them to meet the budget.
SP uses $b$=8 or 16, $c$=64, and varies $\mu$, $\eta$, and query term pruning $\beta$.
ASC is configured with 4096 clusters and 8 segments, and varies $\mu$ and $\eta$. 
BMP uses $b$=8 for $k$=1000 and $b$=32 for $k$=10, 
and varies threshold overestimation ($\alpha$) and query pruning ($\beta$).
Seismic uses a posting list pruning ($\beta$) of 25,000, summary mass ($\alpha$) of 0.4,
query pruning $q\_cut$=10,
and varies the threshold overestimation ratio.
Seismic cannot be rank-safe because it uses static index pruning.

For rank-safe search on SPLADE,
SP is 32\% faster than BMP for $k=10$, and 25\% faster for $k=1000$.
Compared to ASC, SP is about 3.3x faster for both $k$=10 and 1000.
For a recall budget of 99\%,
SP is up to 2.9x faster than BMP, 3.3x faster than Seismic, and 9.1x faster than ASC.
For a recall budget below 99\%,
Seismic is more competitive because
it uses aggressive static index pruning
whereas SP, ASC, and BMP operate on the full index.

\begin{figure}[t!]%
    \begin{center}
      \includegraphics[width=\columnwidth,height=0.5\columnwidth]{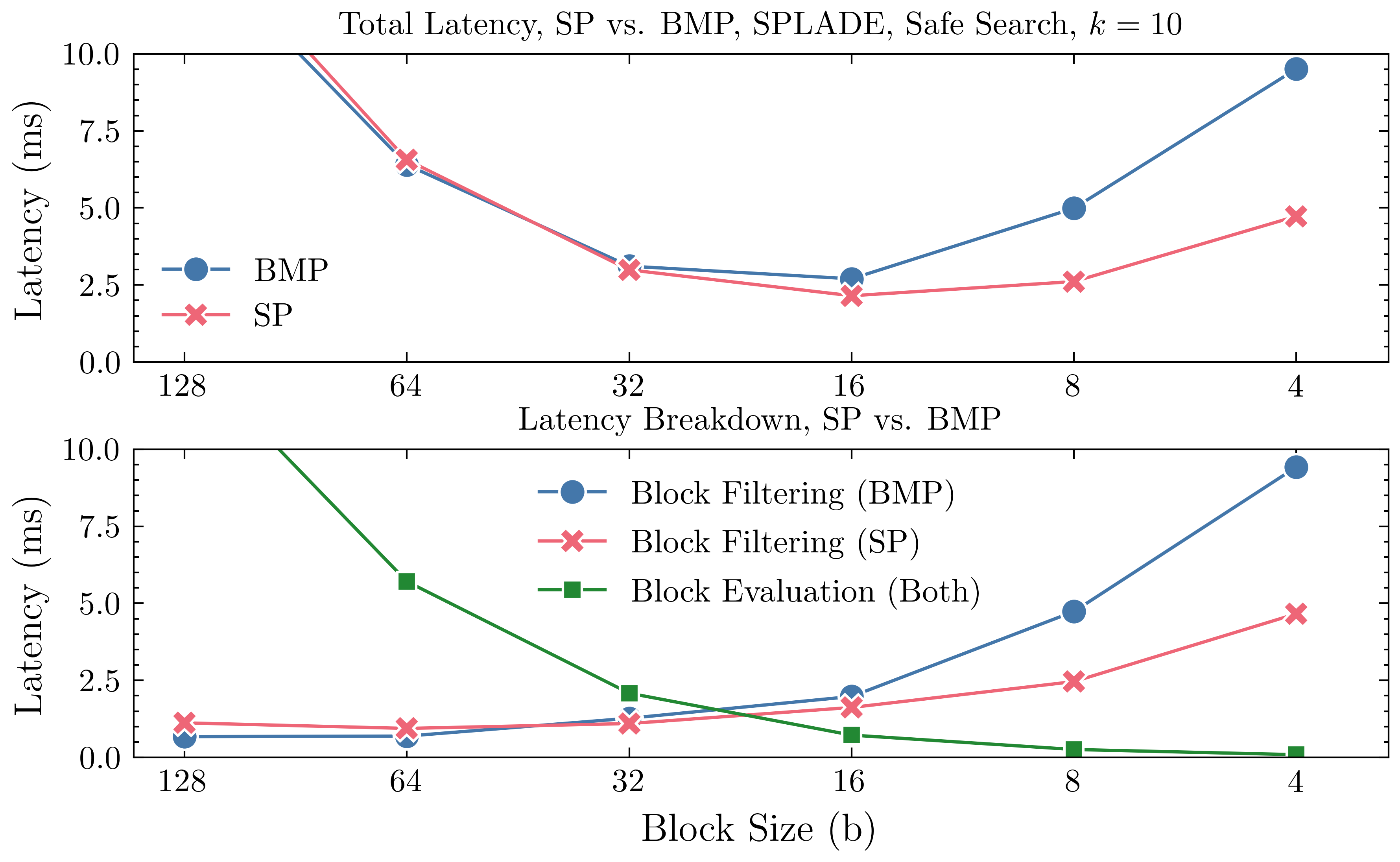}
    \end{center}
    \vspace{-3mm}
      \setlength{\belowcaptionskip}{-5pt}
      \caption{Top: Total latency of SP and BMP when varying  $b$ under safe pruning. 
Cost breakdown in block and superblock filtering, and in document scoring of each un-pruned block}  
    \label{fig:overhead}
\end{figure}

\noindent
{\bf SP vs. BMP with different block size $b$.}
Figure~\ref{fig:overhead} shows the total latency (top) and cost breakdown (bottom)
of SP and BMP
when $b$  decreases from 128 to 8.
When $b$ becomes smaller, BMP achieves a tighter $BoundSum$, but block filtering overhead 
increases.
SP maintains the advantages of evaluating small blocks while reducing the overhead in block and 
superblock filtering.

\noindent
{\bf Effects of superblock  pruning}.
Table \ref{tab:pruning_example}
shows the effectiveness of superblock pruning in SP on SPLADE with $k=10$. 
\#SuB represents the average number of superblocks pruned as a percentage.
\#Bl represents the average number of blocks pruned as a percentage.
\#Bsc represents the average number of un-pruned blocks whose documents are scored. 
MRR is MRR@10. 
Re is Recall@10 for $k=10$ and Recall@1000 for $k=1000$. 
Even for safe search ($\mu=1$), SP is able to prune 24\% of superblocks for $k$=10.
As $\mu$ decreases, the amount of pruning at the superblock level increases significantly.
However, the number of blocks pruned is roughly the same.
This is because the blocks form a tight bound for documents within it;
SP is able to avoid groups of blocks that are unlikely to have any relevant documents,
thus reducing the overhead from computing block $BoundSums$.
Under 100\% probabilistic safeness ($\eta=1$), 
even at $\mu$=0.4 for $k$=1000, there is a negligible impact on 
the relevance metric for the Dev set, 
DL 19, and DL 20, though recall begins to drop when $\mu$=0.4.
In comparison, even a low overestimation threshold in BMP leads to a large drop in relevance.
For overestimation of 0.8 in the same setting, BMP's relevance drops to 62.6 (93.4\% of safe).

\begin{table}[t!]
    \small
    \centering
\caption{Effects of superblock  pruning on SPLADE with $k=10$ when $\mu$ varies. $\eta$=1, $c$=64, $b$=8,
$N \approx 1.1M$}
        
    \label{tab:pruning_example}
    
    \resizebox{0.86\columnwidth}{!}{
    \begin{tabular}{l|ccccc|cc|cc}
    \hline
    \hline
        & \multicolumn{5}{c|}{MS MARCO Dev} & \multicolumn{2}{c|}{DL 19} & \multicolumn{2}{c|}{DL 20}\\ 
    \hline
    \textbf{$\mu$} & \textbf{\#SuB} & \textbf{\#Bl} & \textbf{\#Bsc} & \textbf{MRR} & \textbf{Re} & \textbf{nDCG} & \textbf{Re} & \textbf{nDCG} & \textbf{Re} \\
    \hline
    & \multicolumn{9}{c}{$k$=10} \\ 
    \hline
    1.0 & 24.2\% & 96.6\% & 141 & 38.11 & 66.99 & 73.16 & 17.25 & 71.97 & 24.54 \\
    0.8 & 33.7\% & 96.6\% & 139 & 38.09 & 66.96 & 73.16 & 17.25 & 71.97 & 24.54 \\
    0.6 & 49.5\% & 96.6\% & 139 & 38.09 & 66.96 & 73.16 & 17.25 & 71.97 & 24.54 \\
    0.4 & 74.9\% & 97.0\% & 139 & 38.08 & 66.96 & 73.16 & 17.25 & 71.97 & 24.54 \\
    \hline
    & \multicolumn{9}{c}{$k$=1000} \\
    \hline
    1.0 & 15.7\% & 93.3\% & 4517 & 38.11 & 98.36 & 73.16 & 82.91 & 71.97 & 83.91 \\
    0.8 & 22.1\% & 93.3\% & 4513 & 38.09 & 98.32 & 73.16 & 82.91 & 71.97 & 83.91 \\
    0.6 & 33.8\% & 93.4\% & 4513 & 38.09 & 98.32 & 73.16 & 82.91 & 71.97 & 83.84 \\
    0.4 & 57.0\% & 93.7\% & 4491 & 38.09 & 98.29 & 73.16 & 82.99& 71.97 & 83.84 \\
    \hline
    \hline
    \end{tabular}
    }
\vspace{-5pt}
\end{table}

{\bf CPU cache, superblock size, and overestimation ablation.}
Table \ref{tab:ablation} shows the impact of our cache design for $BoundSum$ computation,
the choice of superblock size,
and the impact of threshold overestimation ($\mu$) when $\eta=1$.
The left side shows latency for our cache-optimized loop with superblock-at-a-time (SaaT) order,
while the right shows 
latency of term-at-a-time (TaaT) order.
SaaT is almost always much faster than TaaT, 
and up to 1.89 times faster.

For small values of $c$,
there are more superblocks,
introducing more superblock-level pruning overhead.
However,
this also permits more accurate pruning at the superblock level
when threshold overestimation increases.
When $\mu$ is 0.6 or higher with superblock-at-a-time order,
$c$=64 or 128 is the best.
When $\mu$ is 0.4, superblock pruning yields more block level pruning,
and $c$=32 is the best.

\begin{table}[t!]
    \small
        \centering
        \caption{Mean response time ($ms$) for two $BoundSum$ computation order options,
        superblock size ($c$), and threshold overestimation ($\mu)$. SPLADE, $k$=10, $\eta$=1, $b$=8}
            
    \label{tab:ablation}
    \resizebox{0.8\columnwidth}{!}{
        \begin{tabular}{l|cccc|cccc}
        \hline
        \hline
        & \multicolumn{4}{c|}{Superblock-at-a-Time Order} & \multicolumn{4}{c}{Term-at-a-Time Order} \\
        \hline
        $\mu$ & $c$=16 & 32 & 64 & 128 & $c$=16 & 32 & 64 & 128 \\
        \hline
        1.0 & 3.05 & 3.24 & 2.58 & 2.52 & 4.03 & 3.92 & 4.02 & 4.10 \\
        0.8 & 2.90 & 2.74 & 2.51 & 2.49 & 3.54 & 4.13 & 4.48 & 4.71 \\
        0.6 & 2.72 & 2.47 & 2.33 & 2.34 & 2.74 & 2.87 & 3.29 & 3.64 \\
        0.4 & 1.78 & 1.68 & 1.76 & 1.93 & 1.72 & 1.77 & 2.20 & 2.76 \\
        \hline
        \hline

    \end{tabular}
    }
\vspace{-10pt}
\end{table}

\noindent
{\bf Comparison on E-SPLADE.}
Table~\ref{tab:esplade}
compares SP with BMP and Seismic under different high-relevance recall budgets on E-SPLADE for $k$=10.
The configuration setting of these algorithms is similar to that for SPLADE.
SP outperforms the other baselines, and is up to 16x faster than Seismic and 1.4x faster than BMP.

\begin{table}[htbp]
        \small
            \centering
            \caption{Mean response time ($ms$) at a fixed Recall@$k$ budget for E-SPLADE $k$=10}
            \renewcommand{\scriptsize}{\small}
        \label{tab:esplade}
        \resizebox{0.99\columnwidth}{!}{
            \begin{tabular}{l|cccccccc}
            \hline
            \hline
            \textbf{Recall} & \multicolumn{2}{c}{\textbf{99\%}} & \multicolumn{2}{c}{\textbf{99.5\%}} & \multicolumn{2}{c}{\textbf{99.9\%}} & \multicolumn{2}{c}{\textbf{Rank-Safe}} \\
            \textbf{Budget} & MRT & MRR & MRT & MRR & MRT & MRR & MRT & MRR \\ 
            \hline
            MaxScore & -- & -- & -- & -- & -- & -- & 8.06 {\scriptsize (15x)} & 38.8 \\
            Seismic & 2.25 {\scriptsize (5.7x)} & 38.6 & 3.06 {\scriptsize (7.1x)} & 38.8 & 7.43 {\scriptsize (16x)} & 38.8 & -- & -- \\
            BMP & 0.476 {\scriptsize (1.2x)} & 38.6 & 0.529 {\scriptsize (1.2x)} & 38.7 & 0.575 {\scriptsize (1.3x)} & 38.8  & 0.723 {\scriptsize (1.4x)} & 38.8 \\ %
            SP & \textbf{0.394} & 38.6 & \textbf{0.430} & 38.7 & \textbf{0.459} & 38.8 &  \textbf{0.530} & 38.8 \\
            \hline
            \hline
    
        \end{tabular}
        }
    \end{table}

\vspace{-3mm}
\section{Conclusion}
We introduced SP,
a novel dynamic pruning scheme
that prunes at the superblock level in addition to the standard block level
 and is designed to exploit CPU cache locality.
Our evaluation demonstrates 
under recall budgets ranging from 99\% or higher on SPLADE,
SP is 2.3x to 3.8x faster than Seismic,
3.2x to 9.4x faster than ASC,
and up to 2.9x faster than BMP on MS MARCO. 
For safe search, SP is up to 1.3x faster than BMP.
SP is  suitable  for speeding up applications that desire high 
relevance.  For such  applications, we recommend setting
$\eta$ close to  1.0, and vary $\mu$ from 0.4 to 1.
Retrieval is a critical component of large-scale search systems and retrieval-augmented generation 
with LLMs (e.g.  ~\cite{NEURIPS2020_RAG,gao2024RAGsurvey,llama2,openai2024gpt4technicalreport}), 
and fast retrieval on low-cost CPUs with high relevance can have a positive impact.

Compared to BMP~\cite{mallia2024BMP},
SP exploits its superblock structure to quickly skip a large number of blocks,
while providing an extra safeguard with $\eta$ for probabilistic safeness.
Compared to Anytime Ranking~\cite{2022ACMTransAnytime}, 
ASC~\cite{qiao2024ASC}, and Seismic~\cite{2024SIGIR-SparseApproximate}, 
SP can handle a much larger number of blocks and overcome the additional overhead
through cache-optimized superblock pruning,
which naturally leads to a tighter bound estimation. 
Seismic~\cite{2024SIGIR-SparseApproximate}  
and Seismic-Wave~\cite{2024CIKM-SeismicWave} 
exploit static index pruning, custom summaries,
and a document proximity graph,
and we hope to explore such techniques in future work.
We will also investigate 
index compression schemes with SP.

{\bf Acknowledgments}.
We thank anonymous referees for their valuable comments.
This work is supported in part by U.S. NSF IIS-2225942
and ACCESS program.
Any opinions, findings, conclusions or recommendations expressed in this material
are those of the authors and do not necessarily reflect the views of the U.S. NSF.

\balance
\newpage
\normalsize



\end{document}